\documentclass[aip, jcp, reprint, a4paper]{revtex4-2}
\usepackage{lmodern}
\usepackage{amsfonts,amssymb,amsmath,bm,graphicx,newtxtext,newtxmath}
\usepackage{hyperref}

\newcommand{\kB}{k_\text{B}}
\newcommand{\LJ}{\text{LJ}}
\newcommand{\LJG}{\text{LJG}}
\newcommand{\rG}{r_{\text{G}}}
\newcommand{\rmd}{\mathrm{d}}
\newcommand{\tilA}{\tilde{A}}
\newcommand{\tG}{\text{G}}
\newcommand{\tpd}{\rho^{(2)}}

\newcommand{\va}{\bm{a}}
\newcommand{\vs}[1]{\bm{#1}}
\newcommand{\vx}{\vs{x}}

\begin{document}
\title{Free-Energy Functional Approach to Inverse Problems for Self-Assembly of Three-Dimensional Crystals}
\author{Masashi Torikai}%
\email{torikai.masashi@mie-u.ac.jp}
\affiliation{Department of Physics Engineering, Faculty of Engineering, Mie University, 1577 Kurimamachiya-cho, Tsu, Mie 514-8507, Japan}

\begin{abstract}
 In this study, a variational method for the inverse problem of self-assembly, i.e., a reconstruction of the interparticle interaction potential of a given structure, is applied to three-dimensional crystals.
 According to the method, the interaction potential is derived as a function that maximizes the free-energy functional of the one- and two-particle density distribution functions.
 The interaction potentials of the target crystals, including those with face-centered cubic (fcc), body-centered cubic (bcc), and simple hexagonal (shx) lattices, are obtained by numerical maximization of the functional.
 Monte Carlo simulations for the systems of particles with these interactions were carried out, and the self-assembly of the target crystals was confirmed for the bcc and shx cases.
 However, in the many-particle system with the predicted interaction for the fcc lattice, the fcc lattice did not spontaneously form and was metastable.
\end{abstract}

\maketitle

\section{Introduction}
The interparticle interaction of a material defines the structure of the material.
Even when the interaction potential is spherically symmetric pairwise, the resulting structures can be complex and diverse.
For example, square, honeycomb, and kagom\'{e} lattices in two-dimensional (2D) space as well as face-centered cubic (fcc), body-centered cubic (bcc), simple hexagonal (shx), hexagonal-close-packed, diamond, and Wurtzite lattices in three-dimensional (3D) space are observed in spherically symmetric pairwise potential systems.
While the bulk one-particle distribution function is uniform in the liquid phase, the pair-distribution function or radial distribution function exhibits various structures.
There exist many methods for determining thermodynamically stable structures of many-particle systems that comprise particles with given interparticle interaction. One example is the classical density functional theory (DFT)\cite{Rascon1996,Rascon1997,Warshavsky2004,Suematsu2012,Suematsu2014}, which determines stable structures by comparing the free energies of several possible structures. There are also a variety of numerical simulation methods\cite{frenkel2001understanding}, including molecular dynamics (MD) and Monte Carlo (MC) methods.

Some problems require the inverse method, which can be used to derive the interparticle interaction with which a many-particle system self-assembles into a given structure.
Inferring the interaction of the components from an experimentally observed macroscopic structure is a typical inverse problem for self-assembly.
In particular, this is useful when it is difficult to measure the interaction directly in experiments.
Another inverse problem is to determine the interparticle interactions according to which model particles spontaneously assemble into an artificially designed structure.
Using the model particles in molecular simulations, we can develop and study model systems with designed structures that have not yet been experimentally synthesized.
The inverse method may also be applicable to the production of functional materials with designed structures made of colloids, whose interaction parameters are experimentally controllable.

Many methods have been proposed for the inverse problem of self-assembly.
For liquids, there are several theoretical methods based on the integral equation approach\cite{Levesque1985,Dharma-wardana1986,Levesque1986,Reatto1986,Rosenfeld1997,Sumi2014a,Sumi2014b,Morita2016,Mashayak2018}.
For solids, the inverse statistical-mechanical method\cite{Rechtsman2005,Rechtsman2006}, relative entropy optimization\cite{Lindquist2016,Lindquist2016b,Jadrich2017,Pineros2017a,Pineros2018,Banerjee2019}, and iterative Boltzmann inversion\cite{Jadrich2015,Lindquist2017} have been proposed.
These inverse methods for solids have been demonstrated to successfully predict the interparticle interactions of various target structures.
These include 2D crystals, such as those with square\cite{Rechtsman2006,Jain2014,Lindquist2016,Pineros2016a,Pineros2017a}, kagom\'{e}\cite{Zhang2013,Lindquist2016,Pineros2016b,Pineros2017a}, honeycomb\cite{Rechtsman2005,Rechtsman2006,Jain2014,Lindquist2016,Pineros2017a}, truncated hexagonal\cite{Pineros2017a,Pineros2017b} and truncated square\cite{Jadrich2017,Pineros2017a,Pineros2017b} lattices. They also include 3D crystals, such as those with simple cubic\cite{Rechtsman2006a}, bcc\cite{Rechtsman2006a}, shx\cite{Rechtsman2006a}, diamond\cite{Rechtsman2007} and Wurtzite lattices\cite{Rechtsman2007}. These methods have also been used to make predictions for other complex structures, such as clusters\cite{Jadrich2015}, porous structures\cite{Lindquist2016b,Lindquist2017}, and periodic crystals in multi-component systems\cite{Pineros2018}.
For solids, the extensive use of molecular simulations is necessary for most of these inverse methods.

In this work, I present an inverse method for self-assembly that does not require molecular simulations to determine interparticle interactions.
The method, called the interaction functional method\cite{Torikai2015}, is a variational method in which the optimal interaction for a given target structure is determined by maximizing the functional of the one- and two-body density distribution functions of the target structure with respect to the interparticle interaction.
In a previous paper\cite{Torikai2015}, I applied the method to 2D target crystals with square, kagom\'{e} and honeycomb lattices.
It was demonstrated via MC simulations that each many-particle system with a predicted interparticle interaction successfully assembled into the given target lattice.
The interaction functional method is thus successful for 2D structures, but it has not yet been examined for 3D structures, which are particularly important for practical applications.
In this paper, by choosing several target structures that are known to self-assemble in the family of interaction potential systems, I examined whether the interaction functional method could correctly predict the interaction potential.
Specifically, I chose bcc, shx, and fcc lattices as target lattices, which were shown to self-assemble in the family of Lennard-Jones-Gauss (LJG) systems via conventional MD study~\cite{Suematsu2012}.
MC simulations for the many-particle systems with predicted interactions demonstrated that the bcc and shx lattices self-assembled, meaning that the method succeeded in determining the correct interaction for these target crystals.
The fcc lattice, however, did not self-assemble in the system with the predicted interaction for the fcc lattice.
Although the fcc lattice was not a stable structure, it was one of the metastable structures in the system.

\section{Method}
\subsection{Formulation of the Interaction Functional Method}
The interaction functional method and its application to 2D crystals were explained in detail in Ref. \onlinecite{Torikai2015}.
Here, I briefly outline the method beginning with the Legendre transformations between two functionals: the grand potential $\Omega[\varphi]$ and intrinsic Helmholtz free energy $A[\rho]$.
Let $\mu$, $\phi_{\text{ext}}(\vx)$, and $\varphi(\vx) = \mu - \phi_{\text{ext}}(\vx)$ be chemical potential, external field, and intrinsic chemical potential, respectively.
Based on the intrinsic Helmholtz free energy of the system, $A[\rho]$, a two-variable functional of $\varphi(\vx)$ and distribution $n(\vx)$ is defined as $\tilde{\Omega}[\varphi, n] = A[n] - \int n(\vx)\varphi(\vx)\rmd \vx$.
The minimum value of $\tilde{\Omega}[\varphi, n]$ with respect to $n(\vx)$ under a given intrinsic chemical potential $\varphi(\vx)$ is the grand potential of the system:
\begin{align}
 \Omega[\varphi] = \inf_{n}\tilde{\Omega}[\varphi, n],
\end{align}
where the grand potential $\Omega[\varphi]$ is defined as
\begin{align}
 \Omega[\varphi] &= -\beta^{-1}\ln \Xi[\varphi],\\
 \Xi[\varphi] &= \sum_{N=0}^{\infty} \int \frac{\rmd \vs{r}^{(N)}}{\lambda^{3N}N!}\exp\biggl[-\beta \sum_{i}^{N - 1}\sum_{j > i}^{N} v(\vs{r}_{i} - \vs{r}_{j}) \notag\\
 &\quad + \beta \sum_{i}^{N}\varphi(\vs{r}_{i})\biggr],
\end{align}
where $\beta$, $\lambda$, $v(\vs{r})$, and $\vs{r}_{i}$ denote the inverse temperature, de Broglie thermal wavelength, interaction potential, and coordinate of the $i$th particle, respectively.
The distribution $n(\vx)$ that gives the minimum of $\tilde{\Omega}[\varphi, n]$ is the one-particle density $\rho(\vx)$ under the intrinsic chemical potential $\varphi(\vx)$:
\begin{align}
 \rho(\vx) = \arg \inf_{n} \tilde{\Omega}[\varphi, n].
\end{align}
The derivation of the density profile $\rho(\vx)$ by the minimization of $\tilde{\Omega}[\varphi, n]$ is a part of DFT\cite{Hansen201361}.
Inversely, from the grand potential $\Omega[\varphi]$ we can define the new functional $\tilA[\rho, \psi] = \Omega[\psi] + \int \rho(\vx)\psi(\vx)\rmd \vx$.
The maximum value of $\tilA[\rho, \psi]$ with respect to $\psi(\vx)$ for a given density profile $\rho(\vx)$ is the intrinsic Helmholtz free energy $A[\rho]$, and the $\psi(\vx)$ that gives the maximum is the intrinsic chemical potential $\varphi(\vx)$\cite{Caillol2002}:
\begin{align}
 A[\rho] = \sup_{\psi}\tilA[\rho, \psi], \quad \varphi(\vx) = \arg \sup_{\psi}\tilA[\rho, \psi]. \label{eq:variational_principle}
\end{align}

The interaction functional method applies an idea of Percus\cite{Percus1962} to this variational principle.
When a particle is fixed at the origin, other particles feel its interaction potential $v(\vx)$ as an external field.
One-particle density with a fixed particle at the origin $\rho_{\text{fix}}(\vx)$ is given by $\rho_{\text{fix}}(\vx) = \tpd(\vx, 0)/\rho(0)$\cite{Percus1962}.
Here, $\rho(0)$ is the one-particle density at the origin, and $\tpd(\vx, 0)$ is the two-particle density distribution function at the origin and the position $\vx$.
Note that both $\rho(0)$ and $\tpd(\vx, 0)$ are the functions in the absence of the fixed particle.
Using Percus's relation to the variational principle \eqref{eq:variational_principle}, we obtain
\begin{align}
 \varphi_{\text{fix}}(\vx) = \arg\sup_{\psi} \tilA\bigl[\tpd/\rho, \psi\bigr],
\end{align}
where $\varphi_{\text{fix}}(\vx)$ is the intrinsic chemical potential in the presence of the particle at the origin, i.e., $\varphi_{\text{fix}}(\vx) = \mu - v(\vx)$.
By maximizing the $\tilA\bigl[\tpd/\rho, \psi\bigr]$ with respect to $\psi(\vx)$, we can determine $\varphi_{\text{fix}}(\vx)$ and consequently the interparticle interaction $v(\vx)$.
The local activity $z(\vx) = \exp[\beta \varphi(\vx)]$ is more useful than the intrinsic chemical potential $\varphi(\vx)$ for the following discussion.

\subsection{Application to Three-Dimensional Crystals}
In implementing the interaction functional method in this study, I used the second-order expansion of the functional $\tilA$ in powers of $\Delta z(\vx) = z_\text{fix}(\vx) - z_{0}$, where $z_\text{fix}(\vx) = \exp[\beta \varphi_\text{fix}(\vx)]$ and $z_{0} = \exp(\beta \mu)$ are the local activities with and without the fixed particle at the origin, respectively.
If we use $\zeta(\vx) = \Delta z(\vx)/z_{0} = \exp[-\beta v(\vx)] - 1$ for the sake of simplifying the notation, the expansion can be expressed as follows:
\begin{align}
 \beta \tilA\bigl[\tpd/\rho, \psi\bigr] &= \beta \Omega[\varphi] + \beta \int\frac{\tpd(\vx, 0)}{\rho(0)}\psi(\vx) \rmd \vx \notag\\
 &\quad - \int\rho(\vx)\zeta(\vx)\rmd \vx - \frac{1}{2}\iint\Bigl[\tpd(\vx, \vx') \notag\\
 &\quad - \rho(\vx)\rho(\vx')\Bigr]\zeta(\vx)\zeta(\vx') \rmd \vx \rmd \vx'. \label{eq:tildeA}
\end{align}
Here, $\rho(\vx)$ and $\tpd(\vx, \vx')$ are the one- and two-particle density distributions of the target crystal, respectively.
As is often the case in classical DFT studies of crystals, the summation of the Gaussians centered at the lattice points
\begin{align}
 \rho(\vx) = \sum_{i}\biggl(\frac{3}{2\pi \sigma_{\rho}^{2}}\biggr)^{3/2}\exp\biggl(-\frac{3}{2\sigma_{\rho}^{2}}(\vx - \va_{i})^{2}\biggr) \label{eq:crystal_density}
\end{align}
was used as the density profile of the target crystal, where $\va_{i}$ denotes the position of the $i$th atom in the target crystal.
The shortest atomic distance (e.g., $a$) is the characteristic length scale in the target crystal.
Hereafter, $a$ is considered to be one unit of the length.
The Lindemann ratio, which is the ratio of the standard deviation $\sigma_{\rho}$ to $a$, was set to $0.15$ in this study, meaning that \eqref{eq:crystal_density} denotes a typical crystal density distribution close to the melting temperature.
For the two-particle distribution function, it was approximated that the $i$th atom fluctuates around $\va_{i}$ independently of the other particles.

As a trial function for the interparticle interaction potential, I used the LJG potential\cite{Rechtsman2006,Engel2007}:
\begin{align}
 v_{\LJG}(r, \rG) &= v_{\text{LJ}}(r) + v_{\text{G}}(r, \rG) \label{eq:LJGpotential}\\
 v_{\LJ}(r) &= \epsilon\Biggl[\biggl(\frac{\sigma}{r}\biggr)^{12} - 2\biggl(\frac{\sigma}{r}\biggr)^{6}\Biggr] \label{eq:LJterm}\\
 v_{\tG}(r, \rG) &= -\epsilon\exp\Biggl[-\frac{(r - \rG)^{2}}{2\xi^{2}\sigma^{2}}\Biggr]. \label{eq:Gterm}
\end{align}
Here, $\sigma$ and $\epsilon$ are constants.
$v_{\LJ}(r)$ is the Lennard-Jones (LJ) potential with the minimum $-\epsilon$ at $r/\sigma = 1$, and $v_{\tG}(r, \rG)$ is the Gaussian well at position $\rG$ with depth $-\epsilon$ and width $\xi\sigma$.
The range of the strong repulsion of the LJ potential, which should be closely related to the shortest atomic distance, is determined by $\sigma$.
In this study, $\sigma$ was chosen to equal to the shortest atomic distance $a$.
The dimensionless constant $\xi$ was set to $\sqrt{0.02}$, as in Ref.~\onlinecite{Suematsu2012}.
The constant $\epsilon$ was chosen so that the minimum of $v_{\LJG}(r)$ would be $-1$.

To be consistent with the choice of the Lindemann ratio, the inverse temperature $\beta$ in \eqref{eq:tildeA} must be set equal to the inverse melting temperature $\beta_{\text{m}}$ of the target crystal.
Although $\beta_{\text{m}}$ cannot be determined before solving the inverse problem, one can estimate that the thermal energy at the melting temperature is of the order of the depth of the potential well.
Therefore, in this study, $\beta$ was set to $1$.
We will return to this point in \S\ref{subsec:bcc_shx}.

LJG systems with $\rG/\sigma=1.0$-$1.4$ have previously been studied using MD simulations by Ref.~\onlinecite{Suematsu2012}.
In this study, self-assembly of fcc, bcc, and shx lattices was observed during the annealing process in systems with $\rG/\sigma = 1.0$, $\rG/\sigma = 1.1$ and $1.2$, and $\rG/\sigma = 1.3$.
However, the system did not crystallize for $\rG/\sigma = 1.4$.
These observations were confirmed by my own MC simulations.

The functional $\tilA\bigl[\tpd/\rho, \psi\bigr]$, the approximation of which is given in \eqref{eq:tildeA}, is a function of $\rG$ because $\zeta(\vx) = \exp[-\beta v_{\LJG}(x, \rG)] - 1$ is used with the trial function \eqref{eq:LJGpotential}.
The next task was to determine the value of the parameter $\rG$ that maximizes $\tilA$ for the one-particle density \eqref{eq:crystal_density} with the lattice points $\{\va_{i}\}$ of a given target crystal.
The integrations in \eqref{eq:tildeA} were performed numerically by Monte Carlo integration using the VEGAS routine in the GNU Scientific Library\cite{GSL}.

\section{Results and Discussion}
The target structures here were the fcc, bcc, and shx lattices, which have previously been observed in the LJG systems currently under consideration\cite{Suematsu2012}.
In what follows, the parameter $\rG$ in \eqref{eq:LJGpotential} was determined for each target crystal using the interaction functional method.
MC simulations were then performed in the $NVT$ ensemble of the LJG system to check whether or not the predicted potential gave rise to the target crystal.
The number of particles $N$ in the simulations was $1500$, and the simulation box was a cube with volume $V = N/\rho$, where $\rho$ is the number density of the target crystal.
Periodic boundary conditions were used in all directions.
A random particle configuration at high temperature was used as the initial configuration in each simulation.
The solid structure obtained after a cooling process was then compared with the target crystal.

\subsection{Body-Centered Cubic and Simple Hexagonal Lattices}\label{subsec:bcc_shx}
\begin{figure}[tbp]
 \centering
 \includegraphics[clip]{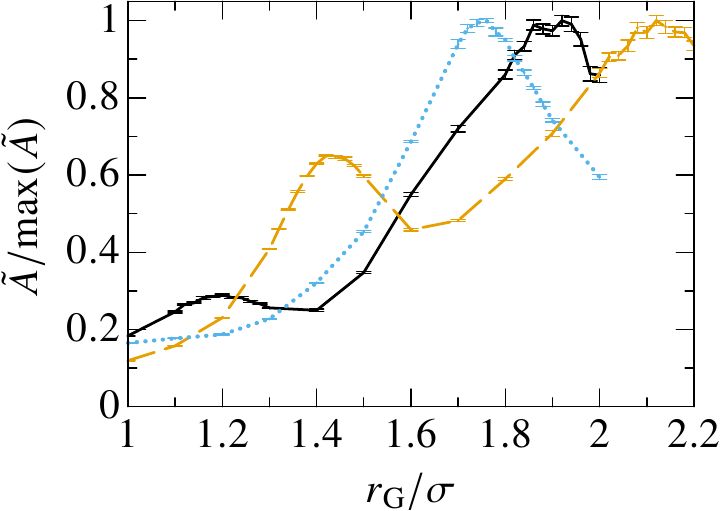}
 \caption{\label{fig:tilA_vs_rG} (Color online) The $\rG$ dependence of the normalized $\tilA$ for the target crystals with bcc (solid black), shx (dashed orange), and fcc lattices(dotted blue). Error bars indicate the estimated error in Monte Carlo integration.}
\end{figure}

\begin{figure}[tbp]
 \centering
 \includegraphics[clip]{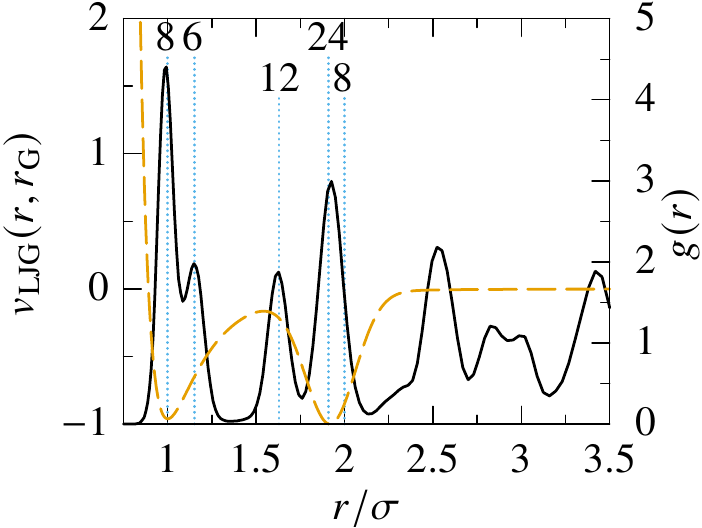}
 \caption{\label{fig:bcc_v_g} (Color online) The LJG potential $v_{\LJG}(r)$ with $\rG/\sigma = 1.92$ (dashed orange) and the radial distribution function $g(r)$ (solid black) for the LJG system obtained using MC simulation. 
 The vertical dotted blue lines and numbers indicate the $n$th nearest neighbor distances and their coordination numbers, respectively, in the bcc target lattice.}
\end{figure}

This section presents the results of the interaction functional method for the bcc and shx target crystals.
The $\rG$ dependence of the functional $\tilA$ for the bcc lattice is shown in Fig.~\ref{fig:tilA_vs_rG}.
The location of the peak, $\rG/\sigma = 1.92$, corresponds to the predicted parameter for the bcc lattice in the interaction functional method framework.
The LJG potential with $\rG/\sigma = 1.92$ is shown in Fig.~\ref{fig:bcc_v_g}.
The value of $\rG$ was out of the range studied by Ref.~\onlinecite{Suematsu2012}, in which bcc was between $\rG/\sigma = 1.1$-$1.2$.

\begin{figure}[tbp]
 \centering
 \includegraphics[clip]{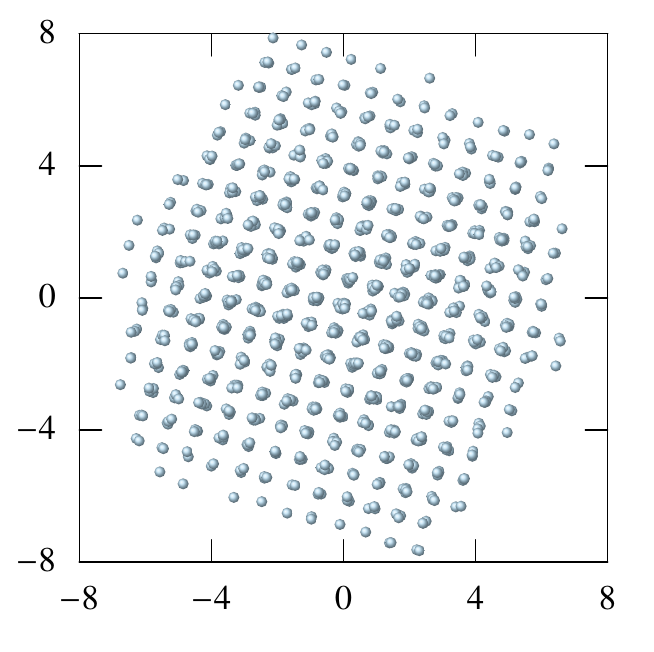}\\
 \includegraphics[clip]{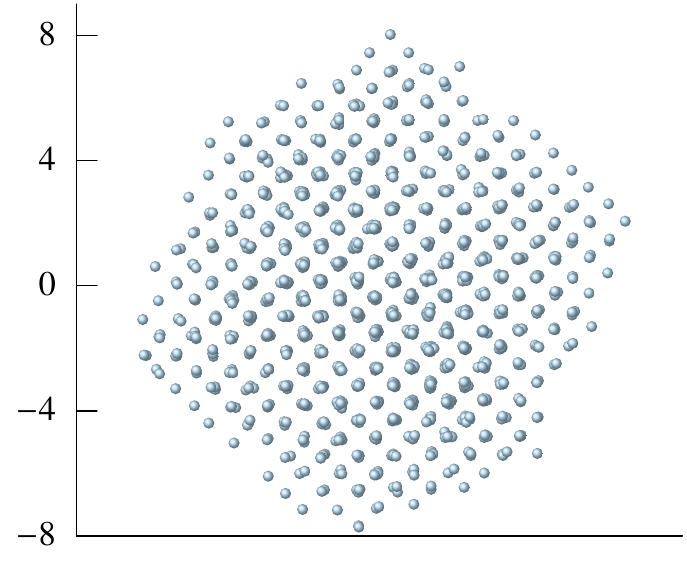}
 \caption{\label{fig:bcc_snapshot} (Color online) Snapshots of the bcc lattice obtained using the MC simulation for the LJG system with $\rG/\sigma = 1.92$, $N = 1500$, and at $\kB T = 1.0$.
 The bcc lattice assembled in the cubic simulation box is rotated so that its four-fold symmetry axis is parallel to the $z$-axis and is viewed along (upper panel) and perpendicular to (lower panel) the $z$-axis.}
\end{figure}

The MC simulation of the LJG system with the predicted $\rG$ revealed that the system successfully self-assembled from the initial random particle configuration into the bcc lattice after the cooling process.
Snapshots of the spontaneously assembled bcc lattice are shown in Fig.~\ref{fig:bcc_snapshot}.
In the snapshots, the overlapping particles along the four-fold symmetry axis indicate a high degree of crystallinity.
The radial distribution function $g(r)$ of the system is shown in Fig.~\ref{fig:bcc_v_g}.
The first to the fifth coordination numbers and nearest neighbor distances of the target bcc lattice are also shown in Fig.~\ref{fig:bcc_v_g}.
The $g(r)$ peaks coincided with the nearest neighbor distances, indicating that the self-assembled structure was in good agreement with the target lattice.

The inverse melting temperature $\beta_{\text{m}}$ of the bcc lattice in the MC simulation was approximately $0.7$.
Using $\beta = 0.7$ in \eqref{eq:tildeA}, the location of the peak in $\tilA$ was $\rG/\sigma = 1.90$.
It was observed that the LJG system with this $\rG$ still self-assembled into the bcc lattice in the MC simulation.
Thus, the prediction of the theory with $\beta = 1$ is almost the same as that with $\beta = \beta_{\text{m}}$.
This was also true for shx and fcc lattices.

\begin{figure}[tbp]
 \centering
 \includegraphics[clip]{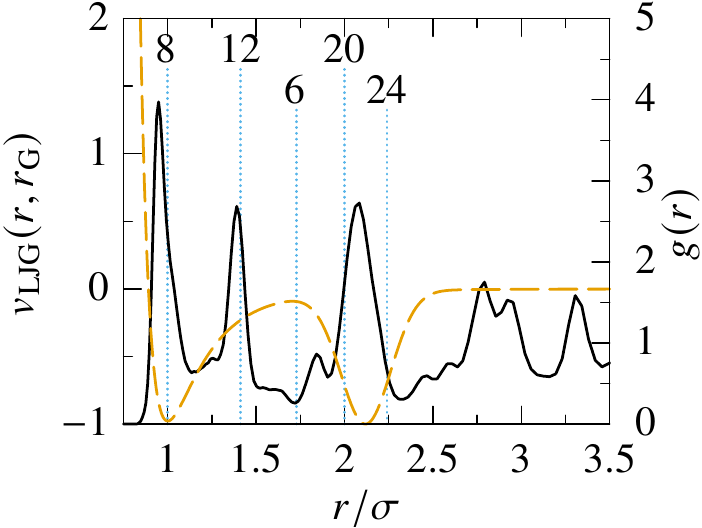}
 \caption{\label{fig:hex_v_g} (Color online) 
The LJG potential $v_{\LJG}(r)$ with $\rG = 2.12$ (dashed orange) and radial distribution function $g(r)$ (solid black) for the LJG system obtained using MC simulation.
 The vertical dotted blue lines and numbers indicate the $n$th nearest neighbor distances and their coordination numbers, respectively, for the target shx lattice.}
\end{figure}

The value of $\tilA$ had a maximum at $\rG/\sigma = 2.12$, when the target structure was a shx lattice (see Fig.~\ref{fig:tilA_vs_rG}).
The LJG potential using this parameter is shown in Fig.~\ref{fig:hex_v_g}.
The solid structure assembled after the cooling process in the MC simulation of the LJG system with $\rG = 2.12$ was a shx lattice, which was verified by examining a snapshot of the structure (not shown).
The $n$th nearest neighbor distance and coordination numbers of the target shx crystal as well as the $g(r)$ obtained in the simulation are shown in Fig.~\ref{fig:hex_v_g}.
Each $g(r)$ peak was at a distance smaller than the corresponding $n$th nearest neighbor distance, indicating that the assembled shx lattice had a smaller lattice constant than that of the target shx lattice.

The functional $\tilA$ for the bcc had a smaller peak at $\rG/\sigma = 1.20$ (Fig.~\ref{fig:tilA_vs_rG}).
Self-assembly of the bcc lattice in the LJG system with this parameter could also be determined using MD\cite{Suematsu2012} and MC simulations.
Although $\tilA$ had a small peak at $\rG/\sigma = 1.42$ for the shx case, the LJG system using this parameter did not assemble into a crystal, let alone the target shx lattice\cite{Suematsu2012}.

To summarize, the predicted LJG systems spontaneously assembled into the target bcc and shx lattices.
These results indicate that the interaction functional method succeeds in predicting the interactions that give rise to these target structures.

\subsection{Face-Centered Cubic Lattice}
\begin{figure}[tbp]
 \centering
 \includegraphics[clip]{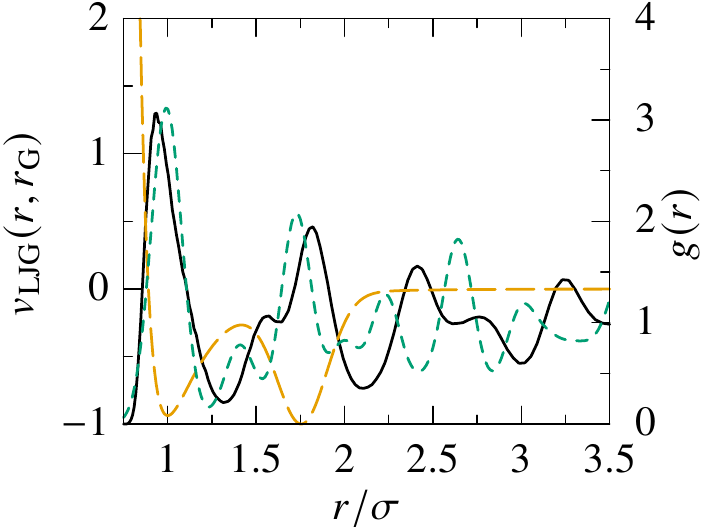}
 \caption{\label{fig:failed_fcc_v_g} (Color online) 
 The LJG potential $v_{\LJG}(r)$ with $\rG = 1.76$ (dashed orange) and the radial distribution function $g(r)$ (solid black) for the LJG system obtained using MC simulation.
 The short dashed green curve represents the radial distribution function of the target fcc lattice, which is calculated from the two-particle density distribution used in the interaction functional method.}
\end{figure}

The functional $\tilA$ had its maximum at $\rG/\sigma = 1.76$ when the fcc lattice was the target structure(see Fig.~\ref{fig:tilA_vs_rG}).
However, it was found that the LJG system with $\rG/\sigma = 1.76$ spontaneously assembled into a bcc lattice instead of an fcc lattice at $\kB T = 2.0$ in the MC simulation.
The $g(r)$ from the simulation is shown in Fig.~\ref{fig:failed_fcc_v_g}.
It can be seen that $g(r)$ in Fig.~\ref{fig:failed_fcc_v_g} has the merged first and second peaks, but that it has nearly identical characteristics to the $g(r)$ of the bcc lattice shown in Fig.~\ref{fig:bcc_v_g}.
The first peak of $g(r)$ was at $r = 0.94\sigma$, indicating that the lattice constant of the bcc lattice was smaller than $\sigma$.
In the LJG system used here, the fcc lattice was stable when $\rG/\sigma = 1.0$, as previously demonstrated in Ref.~\onlinecite{Suematsu2012}.
The $\tilA$ for the fcc lattice (Fig.~\ref{fig:tilA_vs_rG}) did not have a peak at $\rG/\sigma = 1.0$.
Thus, it can be seen that the interaction functional method failed to predict the interaction potential that produces the fcc lattice.

The predicted distance $\rG = 1.76\sigma$ was close to the third nearest neighbor distance of the fcc lattice $1.73\sigma$, whose coordination number $24$ was larger than the first and second coordination numbers $12$ and $6$, respectively.
This fact explains why the fcc $\tilA$ had a peak close to this distance, as the Gauss well at the third nearest neighbor distance was energetically favorable.
The actual structure of the bcc lattice, i.e., that obtained from the simulation results (see Fig.~\ref{fig:failed_fcc_v_g} for its $g(r)$), was energetically stabilized because the first and second nearest neighbor particles were in the LJ potential well, while the third and fourth nearest neighbor particles were in the Gauss well.
The fourth nearest neighbor distance of the bcc lattice had an especially large coordination number of $24$, which considerably reduced the energy of the structure.

Although the target fcc lattice did not self-assemble in the LJG system with the predicted $\rG$, the general stability of the fcc lattice was of interest.
Thus, I performed a heating simulation of the perfect fcc lattice.
The initial perfect fcc lattice had $7 \times 7 \times 7$ conventional unit cells with $N = 1372$ and was put in a cubic simulation box with periodic boundary conditions in all directions.
No phase transitions from the initial fcc lattice to liquid or other crystals were observed for six samples after $5\times 10^{5}$ MC steps at $\kB T = 1.3$ and $1.4$.
However, a phase transition from an fcc to a bcc lattice was found in three samples at $\kB T = 1.5$.
The fact that the fcc lattice persisted at certain temperatures indicates that the target fcc lattice was one of the metastable structures in the predicted LJG system.

\section{Conclusion}
In this study, I applied the interaction functional method to the inverse problem of self-assembly for 3D crystals with bcc, shx, and fcc lattices.
The prediction of the interaction functional method for the LJG potential parameter was obtained for these target crystals.
While MC simulations verified the predictions for the bcc and shx lattices, the predicted LJG system for the fcc lattice was found to self-assemble into a bcc lattice instead of an fcc lattice.

It should be noted that the interaction functional method does not compare the relative stability of the target crystal with other competing structures.
Instead, it simply finds the most favorable interaction potential for the target crystal within the allowed functional forms. 
Therefore, the method cannot exclude competing structures and sometimes fails to predict the correct interaction.
Nevertheless, this method deserves further study, as it has successfully predicted the correct potentials for square, kagom\'{e}, honeycomb, bcc, and shx lattices.

It is also possible that the predictive power of the interaction functional method could be improved despite the inherent shortcoming described above.
It should be noted that the failure to predict the fcc lattice may have resulted from the approximations used in this study.
In particular, the second-order expansion \eqref{eq:tildeA} used in this study is one of the simplest approximations.
Other approximations for the functional that are more sophisticated than functional expansion may improve the prediction, just as the weighted density approximations in classical DFT have led to better results for the problem of freezing\cite{Hansen201361}.
However, a different choice of trial functions is a more feasible approach than finding a better approximation for the functional.
In this paper, the trial function was restricted to the LJG potential characterized by a single parameter $\rG$.
The interaction potential that gives rise to the target structure may be found if a broader functional space is searched.

\begin{acknowledgments}
I would like to thank Professor Akira Yoshimori for his helpful discussions.
 Financial support from the Okasan-Kato Foundation (No. 17-1-26) and the Toyota Physical and Chemical Research Institute (No. 2017-sc-15) is also gratefully acknowledged.
\end{acknowledgments}

%\bibliography{main}

\begin{thebibliography}{38}%
\makeatletter
\providecommand \@ifxundefined [1]{%
 \@ifx{#1\undefined}
}%
\providecommand \@ifnum [1]{%
 \ifnum #1\expandafter \@firstoftwo
 \else \expandafter \@secondoftwo
 \fi
}%
\providecommand \@ifx [1]{%
 \ifx #1\expandafter \@firstoftwo
 \else \expandafter \@secondoftwo
 \fi
}%
\providecommand \natexlab [1]{#1}%
\providecommand \enquote  [1]{``#1''}%
\providecommand \bibnamefont  [1]{#1}%
\providecommand \bibfnamefont [1]{#1}%
\providecommand \citenamefont [1]{#1}%
\providecommand \href@noop [0]{\@secondoftwo}%
\providecommand \href [0]{\begingroup \@sanitize@url \@href}%
\providecommand \@href[1]{\@@startlink{#1}\@@href}%
\providecommand \@@href[1]{\endgroup#1\@@endlink}%
\providecommand \@sanitize@url [0]{\catcode `\\12\catcode `\$12\catcode
  `\&12\catcode `\#12\catcode `\^12\catcode `\_12\catcode `\%12\relax}%
\providecommand \@@startlink[1]{}%
\providecommand \@@endlink[0]{}%
\providecommand \url  [0]{\begingroup\@sanitize@url \@url }%
\providecommand \@url [1]{\endgroup\@href {#1}{\urlprefix }}%
\providecommand \urlprefix  [0]{URL }%
\providecommand \Eprint [0]{\href }%
\providecommand \doibase [0]{https://doi.org/}%
\providecommand \selectlanguage [0]{\@gobble}%
\providecommand \bibinfo  [0]{\@secondoftwo}%
\providecommand \bibfield  [0]{\@secondoftwo}%
\providecommand \translation [1]{[#1]}%
\providecommand \BibitemOpen [0]{}%
\providecommand \bibitemStop [0]{}%
\providecommand \bibitemNoStop [0]{.\EOS\space}%
\providecommand \EOS [0]{\spacefactor3000\relax}%
\providecommand \BibitemShut  [1]{\csname bibitem#1\endcsname}%
\let\auto@bib@innerbib\@empty
%</preamble>
\bibitem [{\citenamefont {Rasc{\'{o}}n}, \citenamefont {Mederos},\ and\
  \citenamefont {Navascu{\'{e}}s}(1996)}]{Rascon1996}%
  \BibitemOpen
  \bibfield  {author} {\bibinfo {author} {\bibfnamefont {C.}~\bibnamefont
  {Rasc{\'{o}}n}}, \bibinfo {author} {\bibfnamefont {L.}~\bibnamefont
  {Mederos}},\ and\ \bibinfo {author} {\bibfnamefont {G.}~\bibnamefont
  {Navascu{\'{e}}s}},\ }\bibfield  {title} {\enquote {\bibinfo {title}
  {{Perturbation Theory for Classical Solids}},}\ }\href
  {https://doi.org/10.1103/PhysRevLett.77.2249} {\bibfield  {journal} {\bibinfo
   {journal} {Physical Review Letters}\ }\textbf {\bibinfo {volume} {77}},\
  \bibinfo {pages} {2249--2252} (\bibinfo {year} {1996})}\BibitemShut {NoStop}%
\bibitem [{\citenamefont {Rasc{\'{o}}n}\ \emph {et~al.}(1997)\citenamefont
  {Rasc{\'{o}}n}, \citenamefont {Velasco}, \citenamefont {Mederos},\ and\
  \citenamefont {Navascu{\'{e}}s}}]{Rascon1997}%
  \BibitemOpen
  \bibfield  {author} {\bibinfo {author} {\bibfnamefont {C.}~\bibnamefont
  {Rasc{\'{o}}n}}, \bibinfo {author} {\bibfnamefont {E.}~\bibnamefont
  {Velasco}}, \bibinfo {author} {\bibfnamefont {L.}~\bibnamefont {Mederos}},\
  and\ \bibinfo {author} {\bibfnamefont {G.}~\bibnamefont {Navascu{\'{e}}s}},\
  }\bibfield  {title} {\enquote {\bibinfo {title} {{Phase diagrams of systems
  of particles interacting via repulsive potentials}},}\ }\href
  {https://doi.org/10.1063/1.473666} {\bibfield  {journal} {\bibinfo  {journal}
  {The Journal of Chemical Physics}\ }\textbf {\bibinfo {volume} {106}},\
  \bibinfo {pages} {6689--6697} (\bibinfo {year} {1997})}\BibitemShut {NoStop}%
\bibitem [{\citenamefont {Warshavsky}\ and\ \citenamefont
  {Song}(2004)}]{Warshavsky2004}%
  \BibitemOpen
  \bibfield  {author} {\bibinfo {author} {\bibfnamefont {V.~B.}\ \bibnamefont
  {Warshavsky}}\ and\ \bibinfo {author} {\bibfnamefont {X.}~\bibnamefont
  {Song}},\ }\bibfield  {title} {\enquote {\bibinfo {title} {{Calculations of
  free energies in liquid and solid phases: Fundamental measure
  density-functional approach}},}\ }\href
  {https://doi.org/10.1103/PhysRevE.69.061113} {\bibfield  {journal} {\bibinfo
  {journal} {Physical Review E}\ }\textbf {\bibinfo {volume} {69}},\ \bibinfo
  {pages} {061113} (\bibinfo {year} {2004})}\BibitemShut {NoStop}%
\bibitem [{\citenamefont {Suematsu}\ \emph {et~al.}(2012)\citenamefont
  {Suematsu}, \citenamefont {Yoshimori}, \citenamefont {Saiki}, \citenamefont
  {Matsui},\ and\ \citenamefont {Odagaki}}]{Suematsu2012}%
  \BibitemOpen
  \bibfield  {author} {\bibinfo {author} {\bibfnamefont {A.}~\bibnamefont
  {Suematsu}}, \bibinfo {author} {\bibfnamefont {A.}~\bibnamefont {Yoshimori}},
  \bibinfo {author} {\bibfnamefont {M.}~\bibnamefont {Saiki}}, \bibinfo
  {author} {\bibfnamefont {J.}~\bibnamefont {Matsui}},\ and\ \bibinfo {author}
  {\bibfnamefont {T.}~\bibnamefont {Odagaki}},\ }\bibfield  {title} {\enquote
  {\bibinfo {title} {{Application of Phase Transition Theory to a Glass-Forming
  System}},}\ }\href {https://doi.org/10.1143/JPSJS.81SA.SA020} {\bibfield
  {journal} {\bibinfo  {journal} {Journal of the Physical Society of Japan}\
  }\textbf {\bibinfo {volume} {81}},\ \bibinfo {pages} {SA020} (\bibinfo {year}
  {2012})}\BibitemShut {NoStop}%
\bibitem [{\citenamefont {Suematsu}\ \emph {et~al.}(2014)\citenamefont
  {Suematsu}, \citenamefont {Yoshimori}, \citenamefont {Saiki}, \citenamefont
  {Matsui},\ and\ \citenamefont {Odagaki}}]{Suematsu2014}%
  \BibitemOpen
  \bibfield  {author} {\bibinfo {author} {\bibfnamefont {A.}~\bibnamefont
  {Suematsu}}, \bibinfo {author} {\bibfnamefont {A.}~\bibnamefont {Yoshimori}},
  \bibinfo {author} {\bibfnamefont {M.}~\bibnamefont {Saiki}}, \bibinfo
  {author} {\bibfnamefont {J.}~\bibnamefont {Matsui}},\ and\ \bibinfo {author}
  {\bibfnamefont {T.}~\bibnamefont {Odagaki}},\ }\bibfield  {title} {\enquote
  {\bibinfo {title} {{Solid phase stability of a double-minimum interaction
  potential system}},}\ }\href {https://doi.org/10.1063/1.4884021} {\bibfield
  {journal} {\bibinfo  {journal} {The Journal of Chemical Physics}\ }\textbf
  {\bibinfo {volume} {140}},\ \bibinfo {pages} {244501} (\bibinfo {year}
  {2014})}\BibitemShut {NoStop}%
\bibitem [{\citenamefont {Frenkel}\ and\ \citenamefont
  {Smit}(2001)}]{frenkel2001understanding}%
  \BibitemOpen
  \bibfield  {author} {\bibinfo {author} {\bibfnamefont {D.}~\bibnamefont
  {Frenkel}}\ and\ \bibinfo {author} {\bibfnamefont {B.}~\bibnamefont {Smit}},\
  }\href {https://books.google.co.jp/books?id=5qTzldS9ROIC} {\emph {\bibinfo
  {title} {Understanding Molecular Simulation: From Algorithms to
  Applications}}},\ \bibinfo {edition} {2nd}\ ed.,\ Computational science
  series\ (\bibinfo  {publisher} {Elsevier Science},\ \bibinfo {year}
  {2001})\BibitemShut {NoStop}%
\bibitem [{\citenamefont {Levesque}, \citenamefont {Weis},\ and\ \citenamefont
  {Reatto}(1985)}]{Levesque1985}%
  \BibitemOpen
  \bibfield  {author} {\bibinfo {author} {\bibfnamefont {D.}~\bibnamefont
  {Levesque}}, \bibinfo {author} {\bibfnamefont {J.~J.}\ \bibnamefont {Weis}},\
  and\ \bibinfo {author} {\bibfnamefont {L.}~\bibnamefont {Reatto}},\
  }\bibfield  {title} {\enquote {\bibinfo {title} {{Pair Interaction from
  Structural Data for Dense Classical Liquids}},}\ }\href
  {https://doi.org/10.1103/PhysRevLett.54.451} {\bibfield  {journal} {\bibinfo
  {journal} {Physical Review Letters}\ }\textbf {\bibinfo {volume} {54}},\
  \bibinfo {pages} {451--454} (\bibinfo {year} {1985})}\BibitemShut {NoStop}%
\bibitem [{\citenamefont {Dharma-wardana}\ and\ \citenamefont
  {Aers}(1986)}]{Dharma-wardana1986}%
  \BibitemOpen
  \bibfield  {author} {\bibinfo {author} {\bibfnamefont {M.~W.~C.}\
  \bibnamefont {Dharma-wardana}}\ and\ \bibinfo {author} {\bibfnamefont
  {G.~C.}\ \bibnamefont {Aers}},\ }\bibfield  {title} {\enquote {\bibinfo
  {title} {{Comment on "Pair Interaction from Structural Data of Dense
  Classical Liquids"}},}\ }\href {https://doi.org/10.1103/PhysRevLett.56.1211}
  {\bibfield  {journal} {\bibinfo  {journal} {Physical Review Letters}\
  }\textbf {\bibinfo {volume} {56}},\ \bibinfo {pages} {1211--1211} (\bibinfo
  {year} {1986})}\BibitemShut {NoStop}%
\bibitem [{\citenamefont {Levesque}, \citenamefont {Weis},\ and\ \citenamefont
  {Reatto}(1986)}]{Levesque1986}%
  \BibitemOpen
  \bibfield  {author} {\bibinfo {author} {\bibfnamefont {D.}~\bibnamefont
  {Levesque}}, \bibinfo {author} {\bibfnamefont {J.~J.}\ \bibnamefont {Weis}},\
  and\ \bibinfo {author} {\bibfnamefont {L.}~\bibnamefont {Reatto}},\
  }\bibfield  {title} {\enquote {\bibinfo {title} {{Levesque, Weis, and Reatto
  Respond}},}\ }\href {https://doi.org/10.1103/PhysRevLett.56.1212} {\bibfield
  {journal} {\bibinfo  {journal} {Physical Review Letters}\ }\textbf {\bibinfo
  {volume} {56}},\ \bibinfo {pages} {1212--1212} (\bibinfo {year}
  {1986})}\BibitemShut {NoStop}%
\bibitem [{\citenamefont {Reatto}, \citenamefont {Levesque},\ and\
  \citenamefont {Weis}(1986)}]{Reatto1986}%
  \BibitemOpen
  \bibfield  {author} {\bibinfo {author} {\bibfnamefont {L.}~\bibnamefont
  {Reatto}}, \bibinfo {author} {\bibfnamefont {D.}~\bibnamefont {Levesque}},\
  and\ \bibinfo {author} {\bibfnamefont {J.~J.}\ \bibnamefont {Weis}},\
  }\bibfield  {title} {\enquote {\bibinfo {title} {{Iterative
  predictor-corrector method for extraction of the pair interaction from
  structural data for dense classical liquids}},}\ }\href
  {https://doi.org/10.1103/PhysRevA.33.3451} {\bibfield  {journal} {\bibinfo
  {journal} {Physical Review A}\ }\textbf {\bibinfo {volume} {33}},\ \bibinfo
  {pages} {3451--3465} (\bibinfo {year} {1986})}\BibitemShut {NoStop}%
\bibitem [{\citenamefont {Rosenfeld}\ and\ \citenamefont
  {Kahl}(1997)}]{Rosenfeld1997}%
  \BibitemOpen
  \bibfield  {author} {\bibinfo {author} {\bibfnamefont {Y.}~\bibnamefont
  {Rosenfeld}}\ and\ \bibinfo {author} {\bibfnamefont {G.}~\bibnamefont
  {Kahl}},\ }\bibfield  {title} {\enquote {\bibinfo {title} {{The inverse
  problem for simple classical liquids: a density functional approach}},}\
  }\href {https://doi.org/10.1088/0953-8984/9/7/004} {\bibfield  {journal}
  {\bibinfo  {journal} {Journal of Physics: Condensed Matter}\ }\textbf
  {\bibinfo {volume} {9}},\ \bibinfo {pages} {L89--L98} (\bibinfo {year}
  {1997})}\BibitemShut {NoStop}%
\bibitem [{\citenamefont {Sumi}\ \emph
  {et~al.}(2014{\natexlab{a}})\citenamefont {Sumi}, \citenamefont {Imamura},
  \citenamefont {Morita}, \citenamefont {Isogai},\ and\ \citenamefont
  {Nishikawa}}]{Sumi2014a}%
  \BibitemOpen
  \bibfield  {author} {\bibinfo {author} {\bibfnamefont {T.}~\bibnamefont
  {Sumi}}, \bibinfo {author} {\bibfnamefont {H.}~\bibnamefont {Imamura}},
  \bibinfo {author} {\bibfnamefont {T.}~\bibnamefont {Morita}}, \bibinfo
  {author} {\bibfnamefont {Y.}~\bibnamefont {Isogai}},\ and\ \bibinfo {author}
  {\bibfnamefont {K.}~\bibnamefont {Nishikawa}},\ }\bibfield  {title} {\enquote
  {\bibinfo {title} {{Model-potential-free analysis of small angle scattering
  of proteins in solution: insights into solvent effects on protein–protein
  interaction}},}\ }\href {https://doi.org/10.1039/C4CP03606A} {\bibfield
  {journal} {\bibinfo  {journal} {Phys. Chem. Chem. Phys.}\ }\textbf {\bibinfo
  {volume} {16}},\ \bibinfo {pages} {25492--25497} (\bibinfo {year}
  {2014}{\natexlab{a}})}\BibitemShut {NoStop}%
\bibitem [{\citenamefont {Sumi}\ \emph
  {et~al.}(2014{\natexlab{b}})\citenamefont {Sumi}, \citenamefont {Imamura},
  \citenamefont {Morita},\ and\ \citenamefont {Nishikawa}}]{Sumi2014b}%
  \BibitemOpen
  \bibfield  {author} {\bibinfo {author} {\bibfnamefont {T.}~\bibnamefont
  {Sumi}}, \bibinfo {author} {\bibfnamefont {H.}~\bibnamefont {Imamura}},
  \bibinfo {author} {\bibfnamefont {T.}~\bibnamefont {Morita}},\ and\ \bibinfo
  {author} {\bibfnamefont {K.}~\bibnamefont {Nishikawa}},\ }\bibfield  {title}
  {\enquote {\bibinfo {title} {{A model-free method for extracting interaction
  potential between protein molecules using small-angle X-ray scattering}},}\
  }\href {https://doi.org/10.1016/j.molliq.2014.03.014} {\bibfield  {journal}
  {\bibinfo  {journal} {Journal of Molecular Liquids}\ }\textbf {\bibinfo
  {volume} {200}},\ \bibinfo {pages} {42--46} (\bibinfo {year}
  {2014}{\natexlab{b}})}\BibitemShut {NoStop}%
\bibitem [{\citenamefont {Morita}\ \emph {et~al.}(2016)\citenamefont {Morita},
  \citenamefont {Uehara}, \citenamefont {Kuwahata}, \citenamefont {Imamura},
  \citenamefont {Shimada}, \citenamefont {Ookubo}, \citenamefont {Fujita},\
  and\ \citenamefont {Sumi}}]{Morita2016}%
  \BibitemOpen
  \bibfield  {author} {\bibinfo {author} {\bibfnamefont {T.}~\bibnamefont
  {Morita}}, \bibinfo {author} {\bibfnamefont {N.}~\bibnamefont {Uehara}},
  \bibinfo {author} {\bibfnamefont {K.}~\bibnamefont {Kuwahata}}, \bibinfo
  {author} {\bibfnamefont {H.}~\bibnamefont {Imamura}}, \bibinfo {author}
  {\bibfnamefont {T.}~\bibnamefont {Shimada}}, \bibinfo {author} {\bibfnamefont
  {K.}~\bibnamefont {Ookubo}}, \bibinfo {author} {\bibfnamefont
  {M.}~\bibnamefont {Fujita}},\ and\ \bibinfo {author} {\bibfnamefont
  {T.}~\bibnamefont {Sumi}},\ }\bibfield  {title} {\enquote {\bibinfo {title}
  {{Interaction Potential between Biological Sensing Nanoparticles Determined
  by Combining Small-Angle X-ray Scattering and Model-Potential-Free Liquid
  Theory}},}\ }\href {https://doi.org/10.1021/acs.jpcc.6b06487} {\bibfield
  {journal} {\bibinfo  {journal} {The Journal of Physical Chemistry C}\
  }\textbf {\bibinfo {volume} {120}},\ \bibinfo {pages} {25564--25571}
  (\bibinfo {year} {2016})}\BibitemShut {NoStop}%
\bibitem [{\citenamefont {Mashayak}, \citenamefont {Miao},\ and\ \citenamefont
  {Aluru}(2018)}]{Mashayak2018}%
  \BibitemOpen
  \bibfield  {author} {\bibinfo {author} {\bibfnamefont {S.~Y.}\ \bibnamefont
  {Mashayak}}, \bibinfo {author} {\bibfnamefont {L.}~\bibnamefont {Miao}},\
  and\ \bibinfo {author} {\bibfnamefont {N.~R.}\ \bibnamefont {Aluru}},\
  }\bibfield  {title} {\enquote {\bibinfo {title} {{Integral equation theory
  based direct and accelerated systematic coarse-graining approaches}},}\
  }\href {https://doi.org/10.1063/1.5020321} {\bibfield  {journal} {\bibinfo
  {journal} {The Journal of Chemical Physics}\ }\textbf {\bibinfo {volume}
  {148}},\ \bibinfo {pages} {214105} (\bibinfo {year} {2018})}\BibitemShut
  {NoStop}%
\bibitem [{\citenamefont {Rechtsman}, \citenamefont {Stillinger},\ and\
  \citenamefont {Torquato}(2005)}]{Rechtsman2005}%
  \BibitemOpen
  \bibfield  {author} {\bibinfo {author} {\bibfnamefont {M.}~\bibnamefont
  {Rechtsman}}, \bibinfo {author} {\bibfnamefont {F.}~\bibnamefont
  {Stillinger}},\ and\ \bibinfo {author} {\bibfnamefont {S.}~\bibnamefont
  {Torquato}},\ }\bibfield  {title} {\enquote {\bibinfo {title} {{Optimized
  Interactions for Targeted Self-Assembly: Application to a Honeycomb
  Lattice}},}\ }\href {https://doi.org/10.1103/PhysRevLett.95.228301}
  {\bibfield  {journal} {\bibinfo  {journal} {Physical Review Letters}\
  }\textbf {\bibinfo {volume} {95}},\ \bibinfo {pages} {228301} (\bibinfo
  {year} {2005})},\ \bibinfo {note} {[Erratum \textbf{97}, 239901
  (2006)]}\BibitemShut {NoStop}%
\bibitem [{\citenamefont {Rechtsman}, \citenamefont {Stillinger},\ and\
  \citenamefont {Torquato}(2006{\natexlab{a}})}]{Rechtsman2006}%
  \BibitemOpen
  \bibfield  {author} {\bibinfo {author} {\bibfnamefont {M.}~\bibnamefont
  {Rechtsman}}, \bibinfo {author} {\bibfnamefont {F.}~\bibnamefont
  {Stillinger}},\ and\ \bibinfo {author} {\bibfnamefont {S.}~\bibnamefont
  {Torquato}},\ }\bibfield  {title} {\enquote {\bibinfo {title} {{Designed
  interaction potentials via inverse methods for self-assembly}},}\ }\href
  {https://doi.org/10.1103/PhysRevE.73.011406} {\bibfield  {journal} {\bibinfo
  {journal} {Physical Review E}\ }\textbf {\bibinfo {volume} {73}},\ \bibinfo
  {pages} {011406} (\bibinfo {year} {2006}{\natexlab{a}})},\ \bibinfo {note}
  {[Erratum \textbf{75}, 019902 (2007)]}\BibitemShut {NoStop}%
\bibitem [{\citenamefont {Lindquist}, \citenamefont {Jadrich},\ and\
  \citenamefont {Truskett}(2016{\natexlab{a}})}]{Lindquist2016}%
  \BibitemOpen
  \bibfield  {author} {\bibinfo {author} {\bibfnamefont {B.~A.}\ \bibnamefont
  {Lindquist}}, \bibinfo {author} {\bibfnamefont {R.~B.}\ \bibnamefont
  {Jadrich}},\ and\ \bibinfo {author} {\bibfnamefont {T.~M.}\ \bibnamefont
  {Truskett}},\ }\bibfield  {title} {\enquote {\bibinfo {title}
  {{Communication: Inverse design for self-assembly via on-the-fly
  optimization}},}\ }\href {https://doi.org/10.1063/1.4962754} {\bibfield
  {journal} {\bibinfo  {journal} {The Journal of Chemical Physics}\ }\textbf
  {\bibinfo {volume} {145}},\ \bibinfo {pages} {111101} (\bibinfo {year}
  {2016}{\natexlab{a}})},\ \Eprint {https://arxiv.org/abs/1609.00851}
  {arXiv:1609.00851} \BibitemShut {NoStop}%
\bibitem [{\citenamefont {Lindquist}, \citenamefont {Jadrich},\ and\
  \citenamefont {Truskett}(2016{\natexlab{b}})}]{Lindquist2016b}%
  \BibitemOpen
  \bibfield  {author} {\bibinfo {author} {\bibfnamefont {B.~A.}\ \bibnamefont
  {Lindquist}}, \bibinfo {author} {\bibfnamefont {R.~B.}\ \bibnamefont
  {Jadrich}},\ and\ \bibinfo {author} {\bibfnamefont {T.~M.}\ \bibnamefont
  {Truskett}},\ }\bibfield  {title} {\enquote {\bibinfo {title} {{Assembly of
  nothing: equilibrium fluids with designed structured porosity}},}\ }\href
  {https://doi.org/10.1039/C5SM03068D} {\bibfield  {journal} {\bibinfo
  {journal} {Soft Matter}\ }\textbf {\bibinfo {volume} {12}},\ \bibinfo {pages}
  {2663--2667} (\bibinfo {year} {2016}{\natexlab{b}})},\ \Eprint
  {https://arxiv.org/abs/arXiv:1603.03408v1} {arXiv:arXiv:1603.03408v1}
  \BibitemShut {NoStop}%
\bibitem [{\citenamefont {Jadrich}, \citenamefont {Lindquist},\ and\
  \citenamefont {Truskett}(2017)}]{Jadrich2017}%
  \BibitemOpen
  \bibfield  {author} {\bibinfo {author} {\bibfnamefont {R.~B.}\ \bibnamefont
  {Jadrich}}, \bibinfo {author} {\bibfnamefont {B.~A.}\ \bibnamefont
  {Lindquist}},\ and\ \bibinfo {author} {\bibfnamefont {T.~M.}\ \bibnamefont
  {Truskett}},\ }\bibfield  {title} {\enquote {\bibinfo {title} {{Probabilistic
  inverse design for self-assembling materials}},}\ }\href
  {https://doi.org/10.1063/1.4981796} {\bibfield  {journal} {\bibinfo
  {journal} {The Journal of Chemical Physics}\ }\textbf {\bibinfo {volume}
  {146}},\ \bibinfo {pages} {184103} (\bibinfo {year} {2017})},\ \Eprint
  {https://arxiv.org/abs/1702.05021} {arXiv:1702.05021} \BibitemShut {NoStop}%
\bibitem [{\citenamefont {Pi{\~{n}}eros}, \citenamefont {Jadrich},\ and\
  \citenamefont {Truskett}(2017)}]{Pineros2017a}%
  \BibitemOpen
  \bibfield  {author} {\bibinfo {author} {\bibfnamefont {W.~D.}\ \bibnamefont
  {Pi{\~{n}}eros}}, \bibinfo {author} {\bibfnamefont {R.~B.}\ \bibnamefont
  {Jadrich}},\ and\ \bibinfo {author} {\bibfnamefont {T.~M.}\ \bibnamefont
  {Truskett}},\ }\bibfield  {title} {\enquote {\bibinfo {title} {{Design of
  two-dimensional particle assemblies using isotropic pair interactions with an
  attractive well}},}\ }\href {https://doi.org/10.1063/1.5005954} {\bibfield
  {journal} {\bibinfo  {journal} {AIP Advances}\ }\textbf {\bibinfo {volume}
  {7}},\ \bibinfo {pages} {115307} (\bibinfo {year} {2017})}\BibitemShut
  {NoStop}%
\bibitem [{\citenamefont {Pi{\~{n}}eros}\ \emph {et~al.}(2018)\citenamefont
  {Pi{\~{n}}eros}, \citenamefont {Lindquist}, \citenamefont {Jadrich},\ and\
  \citenamefont {Truskett}}]{Pineros2018}%
  \BibitemOpen
  \bibfield  {author} {\bibinfo {author} {\bibfnamefont {W.~D.}\ \bibnamefont
  {Pi{\~{n}}eros}}, \bibinfo {author} {\bibfnamefont {B.~A.}\ \bibnamefont
  {Lindquist}}, \bibinfo {author} {\bibfnamefont {R.~B.}\ \bibnamefont
  {Jadrich}},\ and\ \bibinfo {author} {\bibfnamefont {T.~M.}\ \bibnamefont
  {Truskett}},\ }\bibfield  {title} {\enquote {\bibinfo {title} {{Inverse
  design of multicomponent assemblies}},}\ }\href
  {https://doi.org/10.1063/1.5021648} {\bibfield  {journal} {\bibinfo
  {journal} {The Journal of Chemical Physics}\ }\textbf {\bibinfo {volume}
  {148}},\ \bibinfo {pages} {104509} (\bibinfo {year} {2018})},\ \Eprint
  {https://arxiv.org/abs/1801.02750} {arXiv:1801.02750} \BibitemShut {NoStop}%
\bibitem [{\citenamefont {Banerjee}\ \emph {et~al.}(2019)\citenamefont
  {Banerjee}, \citenamefont {Lindquist}, \citenamefont {Jadrich},\ and\
  \citenamefont {Truskett}}]{Banerjee2019}%
  \BibitemOpen
  \bibfield  {author} {\bibinfo {author} {\bibfnamefont {D.}~\bibnamefont
  {Banerjee}}, \bibinfo {author} {\bibfnamefont {B.~A.}\ \bibnamefont
  {Lindquist}}, \bibinfo {author} {\bibfnamefont {R.~B.}\ \bibnamefont
  {Jadrich}},\ and\ \bibinfo {author} {\bibfnamefont {T.~M.}\ \bibnamefont
  {Truskett}},\ }\bibfield  {title} {\enquote {\bibinfo {title} {{Assembly of
  particle strings via isotropic potentials}},}\ }\href
  {https://doi.org/10.1063/1.5088604} {\bibfield  {journal} {\bibinfo
  {journal} {The Journal of Chemical Physics}\ }\textbf {\bibinfo {volume}
  {150}},\ \bibinfo {pages} {124903} (\bibinfo {year} {2019})},\ \Eprint
  {https://arxiv.org/abs/1901.04659} {arXiv:1901.04659} \BibitemShut {NoStop}%
\bibitem [{\citenamefont {Jadrich}\ \emph {et~al.}(2015)\citenamefont
  {Jadrich}, \citenamefont {Bollinger}, \citenamefont {Lindquist},\ and\
  \citenamefont {Truskett}}]{Jadrich2015}%
  \BibitemOpen
  \bibfield  {author} {\bibinfo {author} {\bibfnamefont {R.~B.}\ \bibnamefont
  {Jadrich}}, \bibinfo {author} {\bibfnamefont {J.~A.}\ \bibnamefont
  {Bollinger}}, \bibinfo {author} {\bibfnamefont {B.~A.}\ \bibnamefont
  {Lindquist}},\ and\ \bibinfo {author} {\bibfnamefont {T.~M.}\ \bibnamefont
  {Truskett}},\ }\bibfield  {title} {\enquote {\bibinfo {title} {{Equilibrium
  cluster fluids: Pair interactions via inverse design}},}\ }\href
  {https://doi.org/10.1039/C5SM01832C} {\bibfield  {journal} {\bibinfo
  {journal} {Soft Matter}\ }\textbf {\bibinfo {volume} {11}},\ \bibinfo {pages}
  {9342--9354} (\bibinfo {year} {2015})},\ \Eprint
  {https://arxiv.org/abs/1507.06936} {arXiv:1507.06936} \BibitemShut {NoStop}%
\bibitem [{\citenamefont {Lindquist}\ \emph {et~al.}(2017)\citenamefont
  {Lindquist}, \citenamefont {Dutta}, \citenamefont {Jadrich}, \citenamefont
  {Milliron},\ and\ \citenamefont {Truskett}}]{Lindquist2017}%
  \BibitemOpen
  \bibfield  {author} {\bibinfo {author} {\bibfnamefont {B.~A.}\ \bibnamefont
  {Lindquist}}, \bibinfo {author} {\bibfnamefont {S.}~\bibnamefont {Dutta}},
  \bibinfo {author} {\bibfnamefont {R.~B.}\ \bibnamefont {Jadrich}}, \bibinfo
  {author} {\bibfnamefont {D.~J.}\ \bibnamefont {Milliron}},\ and\ \bibinfo
  {author} {\bibfnamefont {T.~M.}\ \bibnamefont {Truskett}},\ }\bibfield
  {title} {\enquote {\bibinfo {title} {{Interactions and design rules for
  assembly of porous colloidal mesophases}},}\ }\href
  {https://doi.org/10.1039/C6SM02718K} {\bibfield  {journal} {\bibinfo
  {journal} {Soft Matter}\ }\textbf {\bibinfo {volume} {13}},\ \bibinfo {pages}
  {1335--1343} (\bibinfo {year} {2017})},\ \Eprint
  {https://arxiv.org/abs/1612.01880} {arXiv:1612.01880} \BibitemShut {NoStop}%
\bibitem [{\citenamefont {Jain}, \citenamefont {Errington},\ and\ \citenamefont
  {Truskett}(2014)}]{Jain2014}%
  \BibitemOpen
  \bibfield  {author} {\bibinfo {author} {\bibfnamefont {A.}~\bibnamefont
  {Jain}}, \bibinfo {author} {\bibfnamefont {J.~R.}\ \bibnamefont
  {Errington}},\ and\ \bibinfo {author} {\bibfnamefont {T.~M.}\ \bibnamefont
  {Truskett}},\ }\bibfield  {title} {\enquote {\bibinfo {title}
  {{Dimensionality and Design of Isotropic Interactions that Stabilize
  Honeycomb, Square, Simple Cubic, and Diamond Lattices}},}\ }\href
  {https://doi.org/10.1103/PhysRevX.4.031049} {\bibfield  {journal} {\bibinfo
  {journal} {Physical Review X}\ }\textbf {\bibinfo {volume} {4}},\ \bibinfo
  {pages} {031049} (\bibinfo {year} {2014})},\ \bibinfo {note} {[Erratum
  \textbf{4}, 049902 (2014)]},\ \Eprint {https://arxiv.org/abs/1408.6776}
  {arXiv:1408.6776} \BibitemShut {NoStop}%
\bibitem [{\citenamefont {Pi{\~{n}}eros}, \citenamefont {Baldea},\ and\
  \citenamefont {Truskett}(2016{\natexlab{a}})}]{Pineros2016a}%
  \BibitemOpen
  \bibfield  {author} {\bibinfo {author} {\bibfnamefont {W.~D.}\ \bibnamefont
  {Pi{\~{n}}eros}}, \bibinfo {author} {\bibfnamefont {M.}~\bibnamefont
  {Baldea}},\ and\ \bibinfo {author} {\bibfnamefont {T.~M.}\ \bibnamefont
  {Truskett}},\ }\bibfield  {title} {\enquote {\bibinfo {title} {{Breadth
  versus depth: Interactions that stabilize particle assemblies to changes in
  density or temperature}},}\ }\href {https://doi.org/10.1063/1.4942117}
  {\bibfield  {journal} {\bibinfo  {journal} {The Journal of Chemical Physics}\
  }\textbf {\bibinfo {volume} {144}},\ \bibinfo {pages} {084502} (\bibinfo
  {year} {2016}{\natexlab{a}})},\ \Eprint {https://arxiv.org/abs/1603.05989}
  {arXiv:1603.05989} \BibitemShut {NoStop}%
\bibitem [{\citenamefont {Zhang}, \citenamefont {Stillinger},\ and\
  \citenamefont {Torquato}(2013)}]{Zhang2013}%
  \BibitemOpen
  \bibfield  {author} {\bibinfo {author} {\bibfnamefont {G.}~\bibnamefont
  {Zhang}}, \bibinfo {author} {\bibfnamefont {F.~H.}\ \bibnamefont
  {Stillinger}},\ and\ \bibinfo {author} {\bibfnamefont {S.}~\bibnamefont
  {Torquato}},\ }\bibfield  {title} {\enquote {\bibinfo {title} {{Probing the
  limitations of isotropic pair potentials to produce ground-state structural
  extremes via inverse statistical mechanics}},}\ }\href
  {https://doi.org/10.1103/PhysRevE.88.042309} {\bibfield  {journal} {\bibinfo
  {journal} {Physical Review E}\ }\textbf {\bibinfo {volume} {88}},\ \bibinfo
  {pages} {042309} (\bibinfo {year} {2013})}\BibitemShut {NoStop}%
\bibitem [{\citenamefont {Pi{\~{n}}eros}, \citenamefont {Baldea},\ and\
  \citenamefont {Truskett}(2016{\natexlab{b}})}]{Pineros2016b}%
  \BibitemOpen
  \bibfield  {author} {\bibinfo {author} {\bibfnamefont {W.~D.}\ \bibnamefont
  {Pi{\~{n}}eros}}, \bibinfo {author} {\bibfnamefont {M.}~\bibnamefont
  {Baldea}},\ and\ \bibinfo {author} {\bibfnamefont {T.~M.}\ \bibnamefont
  {Truskett}},\ }\bibfield  {title} {\enquote {\bibinfo {title} {{Designing
  convex repulsive pair potentials that favor assembly of kagome and snub
  square lattices}},}\ }\href {https://doi.org/10.1063/1.4960113} {\bibfield
  {journal} {\bibinfo  {journal} {The Journal of Chemical Physics}\ }\textbf
  {\bibinfo {volume} {145}},\ \bibinfo {pages} {054901} (\bibinfo {year}
  {2016}{\natexlab{b}})},\ \Eprint {https://arxiv.org/abs/1607.06528}
  {arXiv:1607.06528} \BibitemShut {NoStop}%
\bibitem [{\citenamefont {Pi{\~{n}}eros}\ and\ \citenamefont
  {Truskett}(2017)}]{Pineros2017b}%
  \BibitemOpen
  \bibfield  {author} {\bibinfo {author} {\bibfnamefont {W.~D.}\ \bibnamefont
  {Pi{\~{n}}eros}}\ and\ \bibinfo {author} {\bibfnamefont {T.~M.}\ \bibnamefont
  {Truskett}},\ }\bibfield  {title} {\enquote {\bibinfo {title} {{Designing
  pairwise interactions that stabilize open crystals: Truncated square and
  truncated hexagonal lattices}},}\ }\href {https://doi.org/10.1063/1.4979715}
  {\bibfield  {journal} {\bibinfo  {journal} {The Journal of Chemical Physics}\
  }\textbf {\bibinfo {volume} {146}},\ \bibinfo {pages} {144501} (\bibinfo
  {year} {2017})},\ \Eprint {https://arxiv.org/abs/1703.08615}
  {arXiv:1703.08615} \BibitemShut {NoStop}%
\bibitem [{\citenamefont {Rechtsman}, \citenamefont {Stillinger},\ and\
  \citenamefont {Torquato}(2006{\natexlab{b}})}]{Rechtsman2006a}%
  \BibitemOpen
  \bibfield  {author} {\bibinfo {author} {\bibfnamefont {M.}~\bibnamefont
  {Rechtsman}}, \bibinfo {author} {\bibfnamefont {F.}~\bibnamefont
  {Stillinger}},\ and\ \bibinfo {author} {\bibfnamefont {S.}~\bibnamefont
  {Torquato}},\ }\bibfield  {title} {\enquote {\bibinfo {title} {{Self-assembly
  of the simple cubic lattice with an isotropic potential}},}\ }\href
  {https://doi.org/10.1103/PhysRevE.74.021404} {\bibfield  {journal} {\bibinfo
  {journal} {Physical Review E}\ }\textbf {\bibinfo {volume} {74}},\ \bibinfo
  {pages} {021404} (\bibinfo {year} {2006}{\natexlab{b}})}\BibitemShut
  {NoStop}%
\bibitem [{\citenamefont {Rechtsman}, \citenamefont {Stillinger},\ and\
  \citenamefont {Torquato}(2007)}]{Rechtsman2007}%
  \BibitemOpen
  \bibfield  {author} {\bibinfo {author} {\bibfnamefont {M.}~\bibnamefont
  {Rechtsman}}, \bibinfo {author} {\bibfnamefont {F.}~\bibnamefont
  {Stillinger}},\ and\ \bibinfo {author} {\bibfnamefont {S.}~\bibnamefont
  {Torquato}},\ }\bibfield  {title} {\enquote {\bibinfo {title} {{Synthetic
  diamond and wurtzite structures self-assemble with isotropic pair
  interactions}},}\ }\href {https://doi.org/10.1103/PhysRevE.75.031403}
  {\bibfield  {journal} {\bibinfo  {journal} {Physical Review E}\ }\textbf
  {\bibinfo {volume} {75}},\ \bibinfo {pages} {031403} (\bibinfo {year}
  {2007})}\BibitemShut {NoStop}%
\bibitem [{\citenamefont {Torikai}(2015)}]{Torikai2015}%
  \BibitemOpen
  \bibfield  {author} {\bibinfo {author} {\bibfnamefont {M.}~\bibnamefont
  {Torikai}},\ }\bibfield  {title} {\enquote {\bibinfo {title} {{Free-energy
  functional method for inverse problem of self assembly}},}\ }\href
  {https://doi.org/10.1063/1.4917175} {\bibfield  {journal} {\bibinfo
  {journal} {The Journal of Chemical Physics}\ }\textbf {\bibinfo {volume}
  {142}},\ \bibinfo {pages} {144102} (\bibinfo {year} {2015})}\BibitemShut
  {NoStop}%
\bibitem [{\citenamefont {Hansen}\ and\ \citenamefont
  {McDonald}(2013)}]{Hansen201361}%
  \BibitemOpen
  \bibfield  {author} {\bibinfo {author} {\bibfnamefont {J.-P.}\ \bibnamefont
  {Hansen}}\ and\ \bibinfo {author} {\bibfnamefont {I.~R.}\ \bibnamefont
  {McDonald}},\ }\href
  {https://doi.org/https://doi.org/10.1016/C2010-0-66723-X} {\emph {\bibinfo
  {title} {Theory of Simple Liquids}}},\ \bibinfo {edition} {4th}\ ed.\
  (\bibinfo  {publisher} {Academic Press},\ \bibinfo {address} {Oxford},\
  \bibinfo {year} {2013})\BibitemShut {NoStop}%
\bibitem [{\citenamefont {Caillol}(2002)}]{Caillol2002}%
  \BibitemOpen
  \bibfield  {author} {\bibinfo {author} {\bibfnamefont {J.-M.}\ \bibnamefont
  {Caillol}},\ }\bibfield  {title} {\enquote {\bibinfo {title} {The density
  functional theory of classical fluids revisited},}\ }\href
  {https://doi.org/10.1088/0305-4470/35/19/301} {\bibfield  {journal} {\bibinfo
   {journal} {Journal of Physics A: Mathematical and General}\ }\textbf
  {\bibinfo {volume} {35}},\ \bibinfo {pages} {4189--4199} (\bibinfo {year}
  {2002})}\BibitemShut {NoStop}%
\bibitem [{\citenamefont {Percus}(1962)}]{Percus1962}%
  \BibitemOpen
  \bibfield  {author} {\bibinfo {author} {\bibfnamefont {J.}~\bibnamefont
  {Percus}},\ }\bibfield  {title} {\enquote {\bibinfo {title} {{Approximation
  Methods in Classical Statistical Mechanics}},}\ }\href
  {https://doi.org/10.1103/PhysRevLett.8.462} {\bibfield  {journal} {\bibinfo
  {journal} {Physical Review Letters}\ }\textbf {\bibinfo {volume} {8}},\
  \bibinfo {pages} {462--463} (\bibinfo {year} {1962})}\BibitemShut {NoStop}%
\bibitem [{\citenamefont {Engel}\ and\ \citenamefont
  {Trebin}(2007)}]{Engel2007}%
  \BibitemOpen
  \bibfield  {author} {\bibinfo {author} {\bibfnamefont {M.}~\bibnamefont
  {Engel}}\ and\ \bibinfo {author} {\bibfnamefont {H.-R.}\ \bibnamefont
  {Trebin}},\ }\bibfield  {title} {\enquote {\bibinfo {title} {{Self-Assembly
  of Monatomic Complex Crystals and Quasicrystals with a Double-Well
  Interaction Potential}},}\ }\href
  {https://doi.org/10.1103/PhysRevLett.98.225505} {\bibfield  {journal}
  {\bibinfo  {journal} {Physical Review Letters}\ }\textbf {\bibinfo {volume}
  {98}},\ \bibinfo {pages} {225505} (\bibinfo {year} {2007})}\BibitemShut
  {NoStop}%
\bibitem [{\citenamefont {Gough}(2009)}]{GSL}%
  \BibitemOpen
  \bibfield  {author} {\bibinfo {author} {\bibfnamefont {B.}~\bibnamefont
  {Gough}},\ }\href@noop {} {\emph {\bibinfo {title} {GNU Scientific Library
  Reference Manual - Third Edition}}},\ \bibinfo {edition} {3rd}\ ed.\
  (\bibinfo  {publisher} {Network Theory Ltd.},\ \bibinfo {year}
  {2009})\BibitemShut {NoStop}%
\end{thebibliography}
%aipnum4-2.bst 2019-01-14 (MD) hand-edited version of apsrev4-1.bst

\end{document}